\documentclass[twocolumn,preprintnumbers,amsmath,amssymb,superscriptaddress]{revtex4}
\usepackage{appendix}
\usepackage{amsmath,amssymb,amsfonts}
\usepackage{graphics}
\usepackage{dcolumn} 
\usepackage{bm,graphicx,color}
\usepackage{float}

\usepackage{braket}

\begin{document}

\title{Magnetic anisotropy and intermediate valence in CeCo$_5$ ferromagnet}

\author{Alexander B. Shick}

\author{Evgenia A. Tereshina-Chitrova}
\affiliation{Institute of Physics, Czech Academy of Sciences, Na Slovance 2, 182 21 Prague,
Czech Republic.}

\newcommand{\orcidauthorA}{0000-0000-000-000X} 




\begin{abstract}
The intermediate valence of Ce in CeCo$_5$ challenges standard density functional theory (DFT) and static DFT+$U$ approaches, which fail to capture its magnetic properties. By combining DFT+$U$ with exact diagonalization of the Anderson impurity model for the Ce 4$f$ shell, we find a substantial reduction of Ce spin and orbital moments, consistent with DFT+DMFT, arising from Ce$^{4+}$ - Ce$^{3+}$ valence fluctuations. The total magnetic moment of 6.70 $\mu_B$ agrees with experiment, and the calculated $4f$ density of states reproduces photoemission and Bremsstrahlung isochromat spectra. The uniaxial magnetic anisotropy energy reaches 4.8 meV/f.u. when Coulomb correlations on both Ce 4$f$ and Co 3$d$ shells are included, in very good agreement with experimental data. These results highlight the importance of dynamical correlations and provide guidance for exploring high-performance, low-rare-earth-content permanent magnets.
\end{abstract}


\date{\today}


\maketitle

\section{Introduction}
The most powerful permanent magnets are typically based on intermetallic compounds that include rare-earth (RE) elements such as Nd, Sm, Dy, and Tb, along with transition metals like Fe and Co~\cite{Gutfleisch2011,Tereshina2013}. Rare-earth elements contribute strong spin-orbit coupling, which is crucial for achieving high magnetocrystalline anisotropy, while transition metals provide high spontaneous magnetization and elevated Curie temperatures.

Due to increasing market demand and reliance on expensive rare-earth elements, there is a growing effort within both industry and academia to develop high-performance magnets with reduced rare-earth content or substituted critical rare earths with more abundant alternatives.
Cerium (Ce) has recently attracted interest for use in rare-earth-transition metal (RE-TM) magnets, owing to its relative abundance compared to more critical rare-earth elements~\cite{Gabay2018,Lamichhane2019}.

The CeCo$_5$ alloy stabilizes in a hexagonal CaCu$_5$-type crystal structure (space group $P6/mmm$) with lattice parameters, a = 4.92~\AA ~ and c = 4.03~\AA, and a preferential direction of magnetization parallel to the hexagonal axis (c-axis)~\cite{Meyer(1987)}. The Co atoms occupy two distinct crystallographic sites: 2c and 3g, while the Ce atoms occupy the 1a site. Beyond its intrinsic properties, the CaCu$_5$-type phase is of broader significance, as it forms the structural building block for more complex RE-Co intermetallic compounds,
including the technologically important RE$_2$Co${17}$ phases. 

Interestingly, CeCo$_5$ deviates from the expected trends in the RECo$_5$ series. Its lattice constants do not follow the typical lanthanide contraction, and its magnetic moment is unusually low compared to neighboring compounds like LaCo$_5$ and PrCo$_5$. Additionally, CeCo$_5$ has a lower Curie temperature than its rare-earth counterparts~\cite{Nordstrom1990}. These anomalies are believed to originate from the mixed-valence nature of cerium, which fluctuates between trivalent (Ce$^{3+}$) and tetravalent (Ce$^{4+}$) states.

Recent work by Xie and Hong~\cite{Xie2022} used density functional theory (DFT) combined with dynamical mean-field theory (DMFT) and the continuous-time quantum Monte Carlo (CTQMC) method~\cite{Haule2007}. Their calculations showed significantly smaller spin and orbital magnetic moments for Ce 4$f$ electrons compared to DFT predictions~\cite{Nordstrom1990, Chouhan2017}, aligning with valence fluctuations between Ce$^{4+}$ and Ce$^{3+}$. However, their study did not cover the magnetic anisotropy energy (MAE).

Experimentally, CeCo$_5$ exhibit a uniaxial magnetic anisotropy energy (MAE) of 10.5 MJ/m$^3$ (5.5 meV/f.u.) in bulk single crystals~\cite{Bartashevich1994} , with reported total magnetic moments between 6.5~$\mu_B$/f.u.~\cite{Andreev1995} and 7.1~$\mu_B$/f.u.~\cite{Bartashevich1994}. Despite the broad success of DFT in describing a wide range of material systems, rare-earth compounds remain a persistent challenge~\cite{Lee2025}, primarily due to the critical role of strong electron-electron interactions. Consequently, it is not surprising that DFT calculations of the MAE in CeCo$_5$ have yielded values of 1.9 meV/f.u.~\cite{Chouhan2017} and 2.0 meV/f.u.~\cite{Nguyen2018}, which are significantly lower than the experimental result~\cite{Bartashevich1994}.
In contrast, another DFT study reported a MAE of 12.3 meV/f.u.~\cite{patrick2019}, which overshoots the experimental range, highlighting inconsistency across theoretical predictions. Attempts to improve MAE results using the DFT+$U$ approach~\cite{AZA1991} have yielded mixed results: one study claimed an improved MAE of 4.75 meV/f.u.~\cite{Lamichhane2019}, while another reported a reduced value of 1.2 meV/f.u.~\cite{Chouhan2017}. The source of this discrepancy remains unclear.

 The remainder of this paper is organized as follows. Section~II briefly describes the DFT+$U$ (ED) methodology. Section~III presents its application to the electronic and magnetic properties of the  CeCo$_5$ferrimagnet, with a comparison to results obtained using other theoretical approaches and to available experimental data. Conclusions are given in Section~IV.
 
\section{Theoretical Method}

In this work, we use the DFT+$U$(ED) extension~\cite{Shick2021,Shick2024} of the DFT+$U$ method that combines the
DFT with the exact diagonalization (ED) of the multiorbital Anderson impurity model (AIM)~\cite{Mahan} Hamiltonian for the 4$f$-shell, and apply 
it to investigate the electronic structure, spin and orbital magnetic moments, and the magnetic anisotropy energy 
in CeCo$_5$. We follow the  "LDA++" methodology~\cite{Lichtenstein1998} and treat explicitly the electron-electron 
interactions in the $4f$-shell using the exact diagonalization of the AIM, whereas  
the $s$, $p$, and $d$ shells are treated in DFT. 

The effects of the Coulomb interaction on the electronic structure are described using an auxiliary impurity model that explicitly treats the full fourteen-orbital Ce 4$f$ shell~\cite{Mahan}. This multi-orbital impurity model incorporates the complete Coulomb interaction, the strong spin-orbit coupling (SOC) characteristic of Ce, and the crystal-field splitting. 
The corresponding Hamiltonian can be expressed as
\begin{align}
\label{eq:hamilt}
H_{\rm int}  = & \sum_{\substack {q m m' \\ \sigma \sigma'}}
 \epsilon^{q}_{m\sigma,m'\sigma'} b^{\dagger}_{qm\sigma}b_{qm'\sigma'}
 +\sum_{m\sigma} \epsilon_f f^{\dagger}_{m \sigma}f_{m \sigma} 
\nonumber \\
& + \sum_{mm'\sigma\sigma'} \bigl(\xi {\bf l}\cdot{\bf s}
  + \Delta_{\rm CF} +  \frac{\Delta_{\rm EX}}{2} \hat{\sigma}_z  \bigr)_{m\sigma,m'\sigma'}
  f_{m \sigma}^{\dagger}f_{m' \sigma'}
\nonumber \\
& +  \sum_{\substack {q m m' \\ \sigma \sigma'}}   \Bigl(
   V^{q}_{m\sigma,m'\sigma'}
 f^{\dagger}_{m\sigma} b_{qm' \sigma'} + \text{h.c.}
  \Bigr)
\\
& + \frac{1}{2} \sum_{\substack {m m' m''\\  m''' \sigma \sigma'}}
  U_{m m' m'' m'''} f^{\dagger}_{m\sigma} f^{\dagger}_{m' \sigma'}
  f_{m'''\sigma'} f_{m'' \sigma}.
\nonumber
\end{align}
Here $f^{\dagger}_{m \sigma}$ creates an electron in orbital $m$
and spin $\sigma$ in the 4$f$ shell
and $b^{\dagger}_{m\sigma}$ creates an electron in state $m\sigma$
in the ``bath'' that
consists of those host band states that hybridize with the impurity
4$f$ shell, 
$\epsilon_f$ is the energy position of the non-interacting 
$4f$ `impurity' level. 
The parameter $\xi$ specifies the strength of
SOC obtained from the atomic potential, and $\Delta_{\rm CF}$ is 
the crystal-field (CF) potential, and $\Delta_{\rm EX}$ is the strength of the exchange field.
An interested reader can find 
further details for a choice of the parameters  in Eq.~(\ref{eq:hamilt}) in the Supplemental information~\cite{supplement}. 

The Lanczos method~\cite{Kolorenc2012} is employed to find
the eigenstates of the many-body Hamiltonian $H_{\rm
imp}$ Eq.(~\ref{eq:hamilt}) and to calculate the one-particle Green's function $[G_{\rm
imp}(z)]_{m m'}^{\sigma \; \; \sigma'}$ in the subspace of the $f$
orbitals. The self-energy
$[\Sigma (z)]_{m m'}^{\sigma \; \; \sigma'}$ is then obtained from
the inverse of the Green's-function matrix $[G_{\rm imp}]$.
Once the self-energy is known, the local Green's function $G(z)$ 
in the subspace of the $f$ spin-orbitals $\{\phi_{\gamma} =  \phi_{m \sigma} \}$,
\begin{equation}
[G(z)]_{\gamma_1 \gamma_2} = \frac{1}{V_{\rm BZ}}
\int_{\rm BZ}{\rm d}^3 k \,\bigl[z+\mu-H_{\rm DFT}({\bf
k})-\Sigma(z)\bigr]^{-1}_{\gamma_1 \gamma_2}\,, \label{eq:gf}
\end{equation}
is calculated in a single-site approximation.~\cite{shick2009}
Then, with the aid of this local Green's function $G(z)$, we evaluate
the occupation matrix $n_{\gamma\gamma'} = -\frac1{\pi}\,\mathop{\rm Im}
\int_{-\infty}^{E_{\rm{F}}} {\rm d} z \, G_{\gamma \gamma'}(z)$.
The matrix $n_{\gamma_1 \gamma_2}$ is used to construct an effective DFT+$U$
potential ${V}_{U}$, which is inserted into the Kohn--Sham-like
equations~\cite{SJDP2004}:
\begin{gather}
\bigl[ -\nabla^{2} + V_{\rm DFT}(\mathbf{r}) + V_{U} + \xi ({\bf l} \cdot
{\bf s}) \bigr]  \Phi_{\bf k}({\bf r}) = \epsilon_{\bf k}^b \Phi_{\bf
k}({\bf r}).
\label{eq:kohn_sham}
\end{gather}
For the spherically-symmetric DFT+$U$  double-counting term (included in  
the potential $V_U$)  we have adopted the fully localized limit
(FLL) form  $V_{dc} = U (n_f-1/2) - J(n_f-1)/2$. We also note that the DFT  potential
$\hat{V}_{\rm DFT}$ in Eq.(\ref{eq:kohn_sham}) acting on the $f$-states 
is corrected to exclude  the non-spherical double-counting 
with $V_U$~\cite{Kristanovski2018}.
 The equations in Eq.~(\ref{eq:kohn_sham}) are iteratively solved until
self-consistency over the charge density is reached.

The updated value of the $n_f$  (4$f$ occupation) determines the corresponding new value of $\epsilon_f$ in Eq.(\ref{eq:hamilt}). The next iteration is then initiated by solving Eq.(\ref{eq:hamilt}) with the updated $\epsilon_f = -V_{dc}$~\cite{shick2009} and computing the corresponding self-energy $\Sigma(z)$. This iterative procedure is repeated until the occupation of the 4$f$ manifold converges to within 0.01. Once the self-consistent DFT+$U$(ED) solution is obtained, the mean-field total energy is evaluated as
$E_{Tot} = E_{\rm DFT} + \Delta E^{ee} $
where $E_{\mathrm{DFT}}$ is the DFT total energy and $\Delta E^{ee} = E^{ee} - E_{dc}$ is the correction term defined as the difference between the electron-electron interaction energy $E^{ee}$ and the double-counting contribution $E_{dc}$ already included in $E_{\mathrm{DFT}}$.

In the DFT and DFT+$U$ calculations, we used  an in-house
implementation~\cite{shick1997, shick99, shick2001} of the full-potential linearized augmented plane-wave (FP-LAPW) method
that includes both scalar-relativistic and spin-orbit coupling effects. The conventional local spin-density approximation (LSDA)
is adopted. The radii of the atomic muffin-tin (MT) spheres are set 
to 2.75~a.u. for the Ce atom, and 2.20~a.u. for Co atoms. The parameter $R_{Co}
\times K_{max}=7.7$ defines the basis set size and the Brillouin zone (BZ) was sampled with 2535 $k$~points.   
The Slater integrals  $F_0=6.73$~eV, and $F_2=8.39$ eV, $F_4=5.61$ eV, and $F_6=$4.15 eV were chosen which correspond to  
Coulomb~$U=6.73$~eV and exchange~$J = 0.70$~eV. These $U$ and $J$ values are in the range of commonly accepted values for
 the rare earths~\cite{vdMarel1988}.
 
 \begin{table*}[!htbp]
\caption{The spin and orbital magnetic moments $\mu_B$, aligned along [001] crystal direction.}
\centering
\begin{tabular}{llcccccccccc}
\hline
\multicolumn{1}{l}{Method} & \multicolumn{2}{c}{Ce-$M^S_{4f}/M^L_{4f}$} & \multicolumn{2}{c}{Ce(1a)-$M^S/M^L$} &\multicolumn{2}{c}{Co(2c)-$M^S/M^L$} &\multicolumn{2}{c}{Co(3g)-$M^S/M^L$} & Total M/f.u.  \\
\hline
\multicolumn{1}{l}{DFT+$U$(ED)} & \multicolumn{2}{c}{-0.14/0.10} & \multicolumn{2}{c}{-0.35/0.12} & \multicolumn{2}{c}{1.28/0.14}& \multicolumn{2}{c}{1.39/0.10} & 6.70 \\
\multicolumn{1}{l}{DFT+DMFT}~\cite{Xie2022} & \multicolumn{2}{c}{-0.10/0.32} &  \multicolumn{2}{c}{-/-} &\multicolumn{2}{c}{1.30/-}& \multicolumn{2}{c}{1.21/-} &  \\
\hline
\multicolumn{1}{l}{DFT(LSDA)} & \multicolumn{2}{c}{-0.53/0.28} & \multicolumn{2}{c}{-0.71/0.29} &\multicolumn{2}{c}{1.31/0.14}& \multicolumn{2}{c}{1.41/0.11} & 6.63 \\
\multicolumn{1}{l}{DFT(GGA)}~\cite{Chouhan2017} & \multicolumn{2}{c}{-/-} &  \multicolumn{2}{c}{-0.73/0.27} &\multicolumn{2}{c}{-/-}& \multicolumn{2}{c}{-/-} & 6.94 \\
\hline
\multicolumn{1}{l}{DFT+$U$}~\cite{Xie2022} & \multicolumn{2}{c}{-0.50/1.12} &  \multicolumn{2}{c}{-/-} &\multicolumn{2}{c}{1.56/0.14}& \multicolumn{2}{c}{1.57/0.11} &  \\
\multicolumn{1}{l}{DFT+$U$}~\cite{Chouhan2017} & \multicolumn{2}{c}{-/-} &  \multicolumn{2}{c}{-0.97/0.57} &\multicolumn{2}{c}{-/-}& \multicolumn{2}{c}{-/-} & 7.23 \\
\hline
\multicolumn{1}{l}{DFT+$U$(HIA)} & \multicolumn{2}{c}{-0.60/2.22} & \multicolumn{2}{c}{-0.81/2.23} & \multicolumn{2}{c}{1.44/0.21}& \multicolumn{2}{c}{1.44/0.11} & 8.97 \\
\hline
\multicolumn{1}{l}{Experiment}    &  && && &&&&\multicolumn{2}{c}{6.5~\cite{Andreev1995},7.1~\cite{Bartashevich1994}}\\
\hline
\end{tabular}
\label{table:1}
\end{table*}

 \section{Results and Discussion}

We now analyze the spin and orbital magnetic moments in CeCo$_5$ with magnetization aligned along the [001] crystallographic axis, as summarized in Table~\ref{table:1}. The spin moment on the Ce(1a) site is antiferromagnetically aligned with those on the Co(2c) and Co(3g) sublattices, in accordance with the Campbell model~\cite{Campbell1972}. The Ce 4$f$-shell spin moment, $M^S_{4f}$, obtained from DFT+$U$(ED) is $-0.14 \mu_B$, in quantitative agreement with recent DFT+DMFT calculations~\cite{Xie2022}. These values are markedly reduced compared to those from itinerant DFT and static mean-field DFT+$U$~\cite{Xie2022}. In the quasi-atomic DFT+$U$(HIA) approach~\cite{shick23}, which neglects hybridization effects in the Anderson impurity model (AIM), $M^S_{4f}$ is slightly enhanced relative to both DFT and DFT+$U$ results.


The Ce 4$f$-shell orbital moment, $M^L_{4f}$, is $0.10 \; \mu_B$ and antiparallel to $M^S_{4f}$, consistent with Hund's third rule. This value agrees reasonably with the DFT+DMFT result of $0.32 \; \mu_B$~\cite{Xie2022}. Both DFT+$U$(ED) and DFT+DMFT yield orbital moments substantially smaller than those obtained from itinerant DFT, DFT+$U$~\cite{Xie2022}, and the localized DFT+$U$(HIA) method.


The suppression of both $M^S_{4f}$ and $M^L_{4f}$ in DFT+$U$(ED) and DFT+DMFT relative to itinerant DFT, static DFT+$U$, and DFT+$U$(HIA) underscores the importance of dynamical electronic correlations, suggesting that hybridization effects play a critical role in moment reduction. This consistent reduction points to an atypical character of the Ce 4$f$ manifold.


The total spin ($M^S$) and orbital ($M^L$) moments within the Ce MT-sphere are also given in Table~\ref{table:1}. The MT-projected spin moment $M^S$ exceeds $M^S_{4f}$ by approximately $0.2  \mu_B$, due to additional contributions from Ce 5$d$ states. In contrast, the orbital moment $M^L$ remains comparable to $M^L_{4f}$ because the 5$d$ states contribute negligibly to orbital magnetism.


For the Co(2c) and Co(3g) sublattices, spin moments from DFT+$U$(ED) are in close agreement with DFT+DMFT results, and in reasonable accord with DFT, DFT+$U$~\cite{Xie2022}, and DFT+$U$(HIA) values. Similarly, $M^L$ values from DFT+$U$(ED) are consistent with DFT and DFT+$U$~\cite{Xie2022}. The DFT+$U$(HIA) method produces a slight enhancement of $M^L$ for the Co(2c) site, while $M^L$ for Co(3g) remains essentially unchanged.


The total magnetic moment per formula unit (f.u.) from DFT+$U$(ED) is 6.70~$\mu_B$, incorporating spin and orbital contributions within the Ce and Co MT-spheres and the interstitial spin polarization. This value lies within the experimental range of 6.5-7.1~$\mu_B$/f.u.~\cite{Andreev1995, Bartashevich1994}, slightly exceeding the LSDA result and falling below the DFT with the generalized gradient approximation (GGA) and GGA+U values reported in Ref.~\cite{Chouhan2017}. In contrast, the localized DFT+$U$(HIA) approach yields a substantially larger total moment, arising from an overestimated Ce $M^L_{4f}$, and falls outside the experimental range.


\begin{figure}[!htbp]
\centerline{\includegraphics[angle=0,width=1.0\columnwidth,clip]{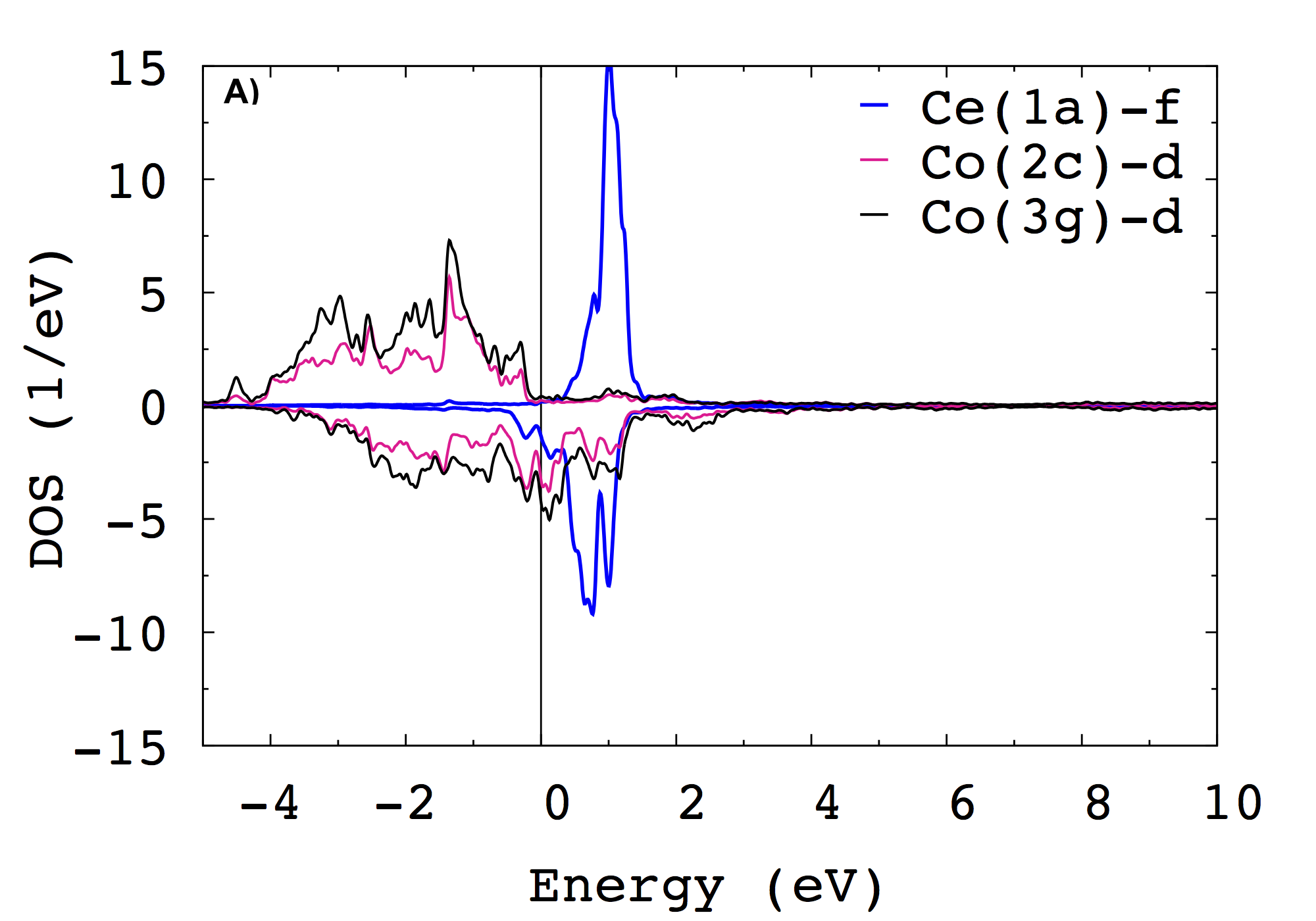}}
\centerline{\includegraphics[angle=0,width=1.0\columnwidth,clip]{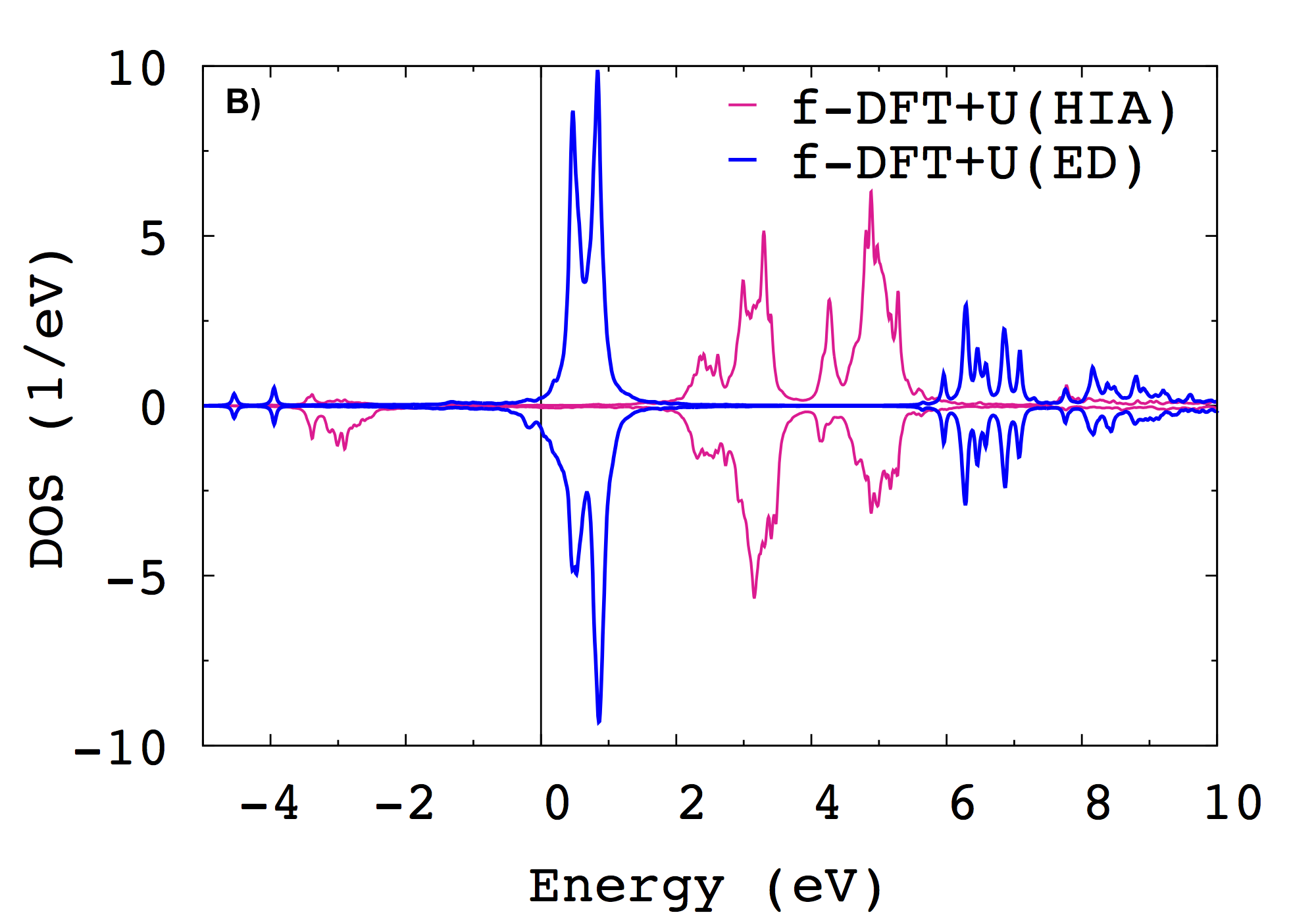}}
\centerline{\includegraphics[angle=0,width=1.0\columnwidth,clip]{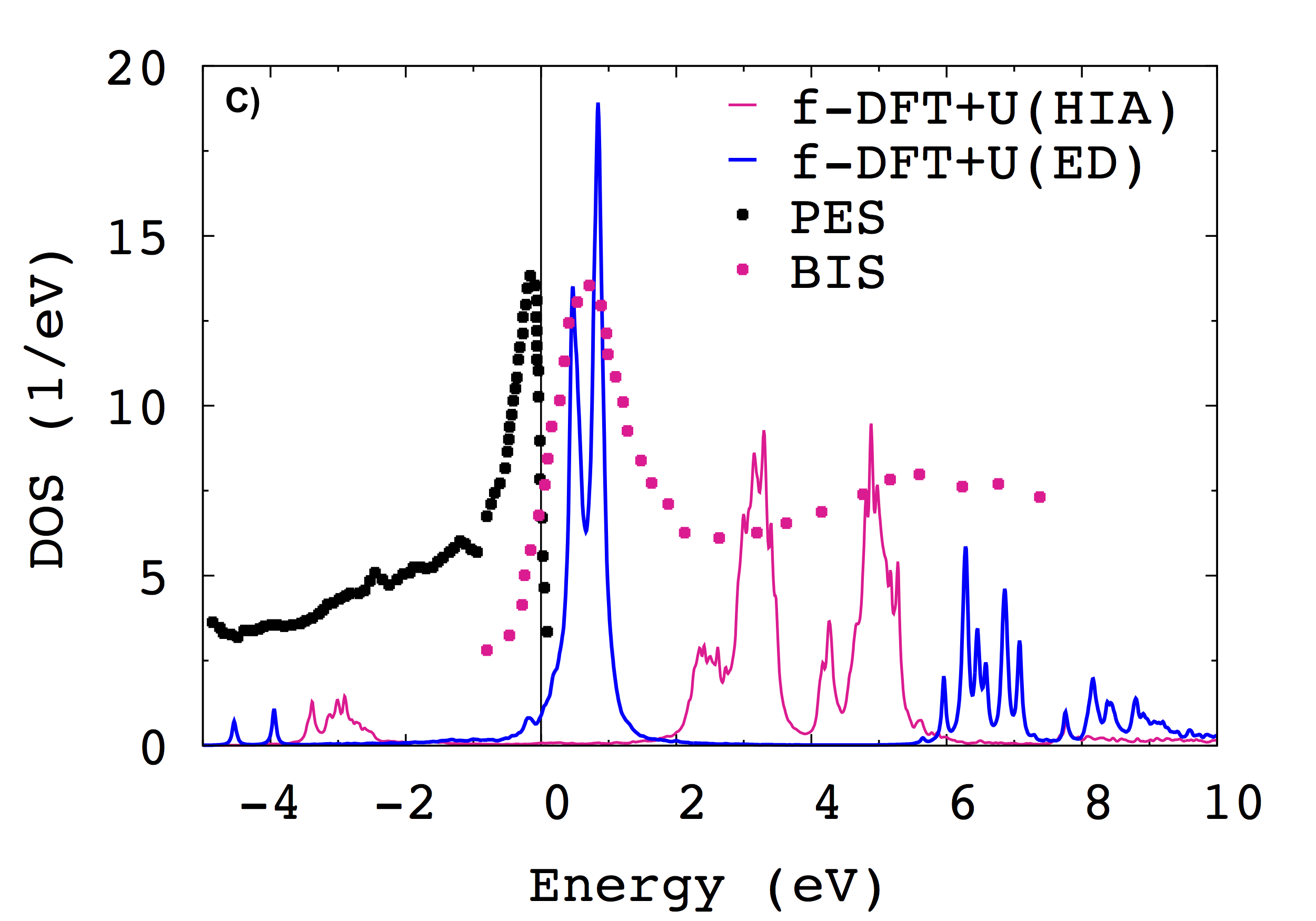}}
\caption{Density of states (DOS) for CeCo$_5$: 
(A) Spin-resolved Ce-$f$ and Co-$d$ projected DOS from DFT(LSDA); 
(B) Spin-resolved Ce-$f$ projected DOS from DFT+$U$(HIA) and DFT+$U$(ED); 
(C) Total Ce-$f$ projected DOS from DFT+$U$(HIA) and DFT+$U$(ED) compared with experimental PES~\cite{imada2001} and BIS~\cite{hillebrecht1984} spectra. 
The experimental PES and BIS spectra are shown in arbitrary units. 
All theoretical spectra are aligned so that the Fermi level is at $E=0$.}
\label{fig:1}
\end{figure}

We computed the density of states (DOS) for CeCo$_5$ to examine its electronic structure. As shown in Fig.~\ref{fig:1}A, the DFT(LSDA) $f$-projected DOS reveals that the $f$-states are positioned near the top of the valence band, while the dominant peaks of the Ce $f$-manifold lie within the conduction band. This distribution is consistent with the tetravalent character of Ce. However, the calculated $n_f$ occupation of 1.01 corresponds to a Ce$^{3+}$ configuration, highlighting the limitations of the itinerant DFT(LSDA) for CeCo$_5$.


The inclusion of the Coulomb-$U$ and exchange-$J$ terms for the Ce 4$f$ shell in a quasi-atomic DFT+$U$(HIA) approximation results in a pronounced splitting between the occupied and unoccupied Ce $4f$ states (Fig.~\ref{fig:1}B). The occupied $4f$ states are positioned approximately 3 eV below the Fermi level ($E_F$), while the unoccupied states are distributed between 2 eV and 5 eV above $E_F$. The calculated $4f$ occupation number, $n_f = 1.00$, corresponds to a trivalent Ce$^{3+}$ configuration. Notably, the Co $3d$ states remain largely unaffected by the application of $U$ and $J$ to the Ce $4f$ shell, indicating weak coupling between the Ce $4f$ and Co $3d$ manifolds.

Incorporating hybridization effects within the DFT+$U$(ED) framework produces a pronounced modification of the $4f$ density of states (DOS), as shown in Fig.~\ref{fig:1}B. The occupied localized peaks are shifted further from $E_F$ and exhibit reduced intensity compared with the DFT+$U$(HIA) results. Additionally, occupied $4f$ states reappear just below $E_F$, resembling the behavior found in LSDA. The unoccupied $4f$ DOS is characterized by two low-energy peaks, located at approximately 0.5 eV and 0.9 eV above $E_F$, along with a broader set of peaks between 5 and 7 eV above $E_F$. These spectral features are in good agreement with those obtained from DFT+DMFT calculations~\cite{Xie2022}. The calculated non-integer occupation, $n_f = 0.64$, indicates an intermediate valence state, consistent with Ce$^{4+}$ -  Ce$^{3+}$ valence fluctuations in CeCo$_5$.
  

A direct validation of the calculated electronic structure can be achieved by comparing the DOS with photoemission (PES) and Bremsstrahlung
 isochromat spectroscopy (BIS) measurements. Fig.~\ref{fig:1}C presents the total $f$-projected DOS obtained from DFT+$U$(HIA) and DFT+$U$(ED) 
 alongside the experimental PES~\cite{imada2001} and BIS~\cite{hillebrecht1984} spectra for CeCo$_5$. Resonant PES data collected near the
  $3d_{5/2} \rightarrow 4f$ transition reveal a pronounced enhancement of the Ce-$4f$ intensity in the energy range from $-1$ to $0$ eV. These
   $f$-states are absent in the DFT+$U$(HIA) calculations but are present in the DFT+$U$(ED) results. The experimental BIS spectrum exhibits two prominent features: a sharp peak at 0.7 eV and a broader peak centered around 5.5 eV above $E_F$. In contrast, the DFT+$U$(HIA) DOS shows pronounced peaks in the  2-4 eV and 4-6 eV energy intervals, neither of which is observed experimentally. The DFT+$U$(ED) results, however, accurately reproduce both major features of the BIS spectrum, highlighting the crucial role of hybridization effects in capturing the correct $f$-state distribution.
   
Within the DFT+$U$(ED) framework, the calculated Ce 4$f$-state occupation is $n_f = 0.64$, corresponding to a singlet many-body ground state (GS) of the Anderson impurity model. The probabilities $P_{f^n}$ of finding the Ce ion in atomic eigenstates with integer 4$f$ occupations are $P_{f^0} = 0.39$, $P_{f^1} = 0.57$, and $P_{f^2} = 0.04$. Thus, the GS is a superposition predominantly of $f^0$ and $f^1$ configurations. The atomic-like $f^1$ state, characterized by $S = 0.5$, $L = 3$, $J = 2.5$, and $M = 2.14 \; \mu_B$, is dynamically screened in CeCo$_5$.
Our DFT+$U$(ED) findings are fully consistent with the DFT+DMFT description~\cite{Xie2022}.


Now, we turn to the salient aspect of our investigation, the magnetic anisotropy properties of  CeCo$_5$.
To the best of our knowledge, a quantitative evaluation of the MAE associated with a dynamically screened magnetic moment has not yet been reported. In the DFT+$U$(ED), the MAE is evaluated as the difference in total energy between 
magnetization along the in-plane [110] and out-of-plane [001] directions: 
MAE =  $E^{[110]} - E^{[001]}$,
where $E^{[110]}$  and  $E^{[001]}$ are the total energies obtained from self-consistent DFT+$U$(ED) calculations 
with the magnetic moments aligned along the [110] and [001] directions, respectively. A positive MAE value indicates that 
the [001] direction is the easy axis of magnetization.

Evaluating the MAE is computationally demanding, as it requires resolving energy differences that are several orders of magnitude smaller than the total energy~\cite{trygg1995}. To ensure consistency, the same k-point mesh is employed in the Brillouin zone (BZ) for both magnetization directions. The BZ integration is performed using the special k-points method. The total energy convergence better than  0.01 meV is preserved in the self-consistent DFT+$U$(ED)  calculations. 
To achieve convergence of the MAE with the desired accuracy, 3825 k-points were used. 
 
The DFT+$U$(ED) calculation yields the MAE value of 2.95 meV/f.u., as shown in Table~\ref{table:2}, which is significantly larger than previous DFT(GGA) results
of 1.9~\cite{Chouhan2017}, and  2.0 meV/f.u.~\cite{Nguyen2018}. For comparison, our DFT (LSDA) calculation yields 2.0 meV/f.u., consistent with these earlier studies. Although the DFT+$U$(ED) value remains below the experimental results~\cite{Bartashevich1994}, it is already substantially enhanced relative to conventional itinerant DFT predictions.

Next, we consider the "localized" quasi-atomic DFT+$U$(HIA) approach, in which the exact atomic limit is preserved. The uniaxial MAE is evaluated using the standard two-sublattice model commonly applied to 3$d$-$f$ intermetallics~\cite{Kuzmin2008}. Within this framework, the Ce 4$f$ crystal-field Hamiltonian is constructed from the self-consistent DFT+$U$(HIA) solution, and the corresponding 4$f$ contribution to the MAE is computed~\cite{yoshioka2018}. The 3$d$ contribution is obtained separately by substituting Ce with Y and evaluating the MAE of YCo$_5$ within DFT.
The resulting Ce-derived MAE contribution of 7.0 meV, combined with the Co-derived contribution of 0.5 meV obtained from the YCo$_5$ MAE calculations, 
yields a total MAE of 7.5 meV (Table~\ref{table:2}), which substantially exceeds the experimental value.\cite{Bartashevich1994}

This two-sublattice approach can also be applied to estimate the Ce-derived contribution to the MAE within the DFT+$U$(ED) framework. By subtracting the DFT-calculated MAE of YCo$_5$ from the total MAE of CeCo$_5$, we obtain a Ce-derived MAE of 2.45 meV, as shown in Table~\ref{table:2}. This reduction reflects the intermediate-valence character of the Ce 4$f$ manifold and the dynamical screening of its magnetic moment. As a result,  the MAE  is significantly reduced relative to the localized {Ce}$^{3+}$ picture.
\begin{table}[!htbp]
\caption{The total and element-specific contributions to the MAE in comparison with the 
experimental data (meV/f.u.)}
\centering
\begin{tabular}{cccccccccc}
\hline
  Method      & Total MAE       & Ce-MAE                 &  Co-MAE \\ 
\hline
{DFT+$U$(ED)} & 2.95               &  2.45                       & 0.50 \\
{DFT+$U$(ED) + } & 4.84               &  2.45                       & 2.39 \\
DFT+$U$ on $d$-Co   &                      \\
\hline
{DFT} & 2.02               &                                &  \\ 
{DFT(GGA)~\cite{Chouhan2017} }   &1.94                &                               &       \\
{DFT(GGA)~\cite{Nguyen2018}}    & 2.04                &                               &       \\
\hline
{DFT+$U$(HIA)} & 7.47                  & 6.97                 & 0.50       \\
\hline
Exp.~\cite{Bartashevich1994}& 5.5 \\
\hline
\end{tabular}
\label{table:2}
\end{table}

So far, our analysis has focused on the influence of the Coulomb interaction $U$ and exchange parameter $J$ on the Ce 4$f$ states. However, these interactions can also affect the Co 3$d$ states.  We then performed DFT+$U$ calculations for YCo$_5$ with $U=1.9$ eV and $J=0.8$ eV~\cite{Nguyen2018}, which yielded a MAE of 2.39 meV. 
Using the two-sublattice approximation, and adding this Co contribution to the Ce-derived MAE gives a corrected total MAE of 4.8 meV (Table~\ref{table:2}). This corrected value of MAE agrees very well with the bulk experimental measurements~\cite{Bartashevich1994}.

\section{Conclusions}
 
To conclude, previous studies of the magnetism and magnetic anisotropy of CeCo$_5$ using DFT and DFT+$U$ approaches did not account for dynamic correlation effects 
and consequently failed to capture the intermediate valence of Ce. The critical role of these correlations has been demonstrated by recent DFT+DMFT calculations~\cite{Xie2022}, 
which report spin and orbital moments of Ce that are substantially smaller than those predicted by itinerant DFT and static mean-field DFT+$U$ methods. 
Nevertheless, due to current technical limitations, DFT+DMFT cannot yet achieve the numerical precision required for quantitatively reliable predictions of the magnetic anisotropy energy in CeCo$_5$.

We have carried out a comprehensive investigation of the electronic structure, magnetic properties, and magnetic anisotropy of CeCo$_5$ using the DFT+$U$(ED) approach, which incorporates exact diagonalization of the Ce 4$f$ shell within the Anderson impurity model. This method significantly reduces the spin and orbital moments of the Ce 4$f$ states compared to standard DFT and static DFT+$U$, in agreement with DFT+DMFT results~\cite{Xie2022}, and underscores the crucial role of dynamical correlations and hybridization effects. The DFT+$U$(ED) framework also accurately reproduces the hybridization-induced modifications of the $4f$ density of states, consistent with experimental photoemission and Bremsstrahlung isochromat spectra and with Ce$^{4+}$-Ce$^{3+}$ valence fluctuations. 

The calculated uniaxial MAE of 2.95 meV/f.u. is significantly enhanced relative to itinerant DFT (2.0 meV/f.u.) and considerably lower than the quasi-atomic DFT+$U$(HIA) result (7.5 meV/f.u.).  Furthermore, inclusion of the Coulomb $U$ and exchange $J$ on the Co 3$d$ shell, combined with with the two-sublattice approximation, yields a corrected MAE of 4.8 meV/f.u.,  that successfully estimates the experimental MAE. Overall, our results provide a quantitatively accurate description of CeCo$_5$ magnetism and offer important insight into the search of high-performance, low-cost rare-earth permanent magnets.

\section{Acknowledgments}
We acknowledge partial support 
provided by the Czech Science Foundation (GACR) Grant
No. 24-11992S. This work was partially co-funded by the European Union and the Czech
Ministry of Education, Youth and Sports (Project TERAFIT -
CZ.02.01.01/00/22$_{-}$008/0004594).

\section{Data availability}
The data that support the findings of this article are openly available~\cite{dataset}.

\end{document}